\documentclass[a4paper]{jpconf}
\usepackage{graphicx}
\usepackage{amssymb,amsfonts}
\usepackage{dsfont}

\newcommand{\half}{\mbox{$\textstyle \frac{1}{2}$}}

\newcommand{\rd}{\mbox{$\rm d$}}

\newtheorem{prop}{Proposition}

\begin{document}

\title{Universal Quantum Measurements} 

\author{Dorje C. Brody${}^{1,2}$ and Lane P. Hughston${}^{1,2}$}

\address{${}^{1}$Department of Mathematics, Brunel University London, Uxbridge UB8 3PH, UK \\ 
${}^{2}$Department of Optical Physics and Modern Natural Science, St Petersburg National Research University of Information Technologies, Mechanics and Optics,\\
49 Kronverksky Avenue, St Petersburg 197101, Russia
}

\ead{Dorje.Brody@brunel.ac.uk, Lane.Hughston@brunel.ac.uk}

\begin{abstract}
We introduce a family  of operations in quantum mechanics that one can regard as ``universal quantum measurements" (UQMs). These measurements are applicable to all finite-dimensional quantum systems and entail the specification of only a minimal amount of structure. The first class of UQM that we consider  involves the specification of the initial state of the system---no further structure is brought into play.  We call operations  of this type ``tomographic measurements", since given the statistics of the outcomes one can deduce the original state of the system. Next, we construct a disentangling operation, the outcome of which, when the procedure is applied to a general mixed state of an entangled composite system, is a disentangled product of pure constituent states. This operation exists whenever the dimension of the Hilbert space is not a prime, and can be used to model the decay of a composite system. As another example, we show how one can make a measurement of the direction along which the spin of a particle of spin $s$ is oriented ($s =  \half, 1, \dots$). The required additional structure in this case involves the embedding of ${\mathds C}{\mathbb P}^{1}$ as a rational curve of degree $2s$ in ${\mathds C}{\mathbb P}^{2s}$. 
\end{abstract}

\section{Introduction}
\label{sec:1} 

As we enter into what may be the dawning of an age of quantum engineering, the widespread interest in quantum 
information, quantum communication, quantum cryptography, and quantum computation thus entailed has had the 
effect of reawakening research in finite dimensional quantum systems. Indeed, from a mathematical point of view an 
$n$-dimensional quantum system can be given a very satisfactory treatment, bringing elements of algebraic geometry, 
symplectic geometry, Riemannian geometry, and convex analysis into play that are perhaps less obviously central in 
the infinite-dimensional situation. Foundational issues are being revisited as well, in part with a view 
to underpinning applications, but also on account of the fact that when the fog of technical issues associated with the 
infinite dimensional case clears one can in the finite-dimensional case---in its own way no less rich---identify more 
distinctly some of the problematic aspects of the general theory. 
 
As a way of honouring the scientific career of Professor Bogdan Mielnik we propose in the present paper to construct a variety of generalised measurement operations arising in the finite-dimensional case that only involve a minimal amount 
of structure on the Hilbert space. To the extent that the structure involved is indeed minimal, the 
class of measurement operations associated with it is maximal---or ``universal''---that is to say, applicable to any 
quantum system. We shall look at a number of explicit examples of such universal quantum measurements (UQMs) and comment on their possible applications. 

We begin in Section~\ref{States and operations} with a brief r\'esum\'e of quantum mechanics from a modern perspective, 
emphasising geometrical and probabilistic aspects of the theory. The idea is to 
model simple quantum systems in such a way that they can be regarded as elements of more complicated systems. 
To model a quantum system we need to specify a collection of suitably inter-related mathematical objects representing 
different aspects of the system. In the spirit of von Neumann 
(1955, IV.1) we can thus say that when a quantum system is given it is ``characterised for the experimenter by the 
enumeration of all the effectively measurable quantities in it and their functional relations with one another." We keep 
the point of view pragmatic, with a view to 
modelling various systems of the type that might be involved in applications to information technology.  No attempt will be made to model the universe as a whole, or to address the 
``measurement problem'', and the models we look at will be mainly nonrelativistic---or, more precisely, pre-relativistic, 
since we do not bring the geometry of space and time into play. 

We find it convenient to use an index notation for Hilbert space operations in our development of the theory of quantum 
state transformations. In Proposition~\ref{Positive maps} we present a useful characterisation of positive maps, and in 
Proposition~\ref{Completely positive maps} we present a simplified proof of Choi's theorem on completely positive maps. 
In Section~\ref{Experiments and state transformations} we use the language of $\sigma$-algebras to present a rather general approach to modelling experiments 
on quantum systems and the state transformations that occur when an experiment is performed. Then in 
Section~\ref{Tomographic measurements} we turn to the construction of  universal quantum measurements that do not 
involve any structure other than that which is implicit in the geometry of the Hilbert space of the system. In particular, no 
preferred ``observables'' are selected. Such a measurement can be regarded as a determination of the state of the 
system. The input state can be any mixed state, and the output is a pure state. If  measurements are performed on a 
large number of independent identical copies of a quantum system, then by gathering the data of the resulting pure 
output states one can determine the input state. Thus, such measurements are ``tomographic'' or  ``informationally complete'' . 

Then we consider examples involving some symmetry breaking, but again without the specification of any observables. 
The first of these is a ``disentangling'' operation, introduced in Section~\ref{Disentangling operations}. These 
operations exist whenever the dimension of the Hilbert 
space is not a prime, and rely on the fact that if the dimension of the Hilbert space is a composite then the 
Hilbert space can be regarded as the tensor product of two or more Hilbert spaces of lower dimension, which we 
regard as ``constituent'' spaces. Disentangling measurements take the form of basic UQMs (of the type described 
above) but operating at the level of the constituents. The input state is a general mixed state, and the output state takes 
the form of a disentangled composite of pure states. The required additional structure involves a canonical Segre embedding of form 
\begin{eqnarray}
{\mathds C}{\mathbb P}^{p_1 -1} \times {\mathds C}{\mathbb P}^{p_2 -1} \times  {\mathds C}{\mathbb P}^{p_3 -1} 
\times \cdots  \,\, \longrightarrow \,\, {\mathds C}{\mathbb P}^{n -1},
\end{eqnarray}
where the prime factorisation of $n$ is given by
\begin{eqnarray}
n = \prod_{i} p_i \, . 
\end{eqnarray}
Finally, in Section~\ref{Coherent measurements} we consider a class of ``coherent'' measurements 
that generalise the idea of determining the axis of spin in the case of a particle with spin.  The additional structure involved in this class of generalised measurements takes the form of an embedding of  ${\mathds C}{\mathbb P}^{n-1}$ as a rational variety of degree $d$ in 
${\mathds C}{\mathbb P}^{N-1}$, where 
\begin{eqnarray}
N = \frac { (n+d-1)!} {  (n-1)! \, (d)! } .
\label{Veronese embedding dimension}
\end{eqnarray}
For example, if $n = 3$, the additional structure is given by the embedding of ${\mathds C}{\mathbb P}^{2}$ as a rational surface of degree $d$ in 
${\mathds C}{\mathbb P}^{N-1}$ for $N =  \half (d+1)(d+2)$. In the case of degree two we obtain the classical  Veronese embedding of  
${\mathds C}{\mathbb P}^{2}$ as a rational surface in ${\mathds C}{\mathbb P}^{5}$.

\section{States and operations} 
\label{States and operations}

In the standard approach to quantum theory taught to undergraduate physicists, the usual line of attack is to introduce states as elements 
of a Hilbert space, then observables as Hermitian operators; a Hamiltonian is introduced as a special observable, and 
the evolution of the state vector, in the absence of external interventions, is given by the Schr\"odinger equation. 
When a measurement of an observable is made, the outcome is one of the eigenvalues of the associated Hermitian 
operator, and the probability of any particular outcome can be worked out using the Born rule; the associated change in the 
state of the system is then determined by the so-called projection postulate of von Neumann in the more general form 
introduced by L\"uders (1951) to take into account degeneracies (see Adler {\it et al.}~2001, sections 2-3,  for a discussion of the status of the L\"uders postulate). We shall assume that the reader is  familiar with the standard approach, and how it is conventionally 
applied, and we take for granted its many strengths.  

Nevertheless, from a modern perspective we can to some extent dispense with the foundational notions of observables as Hermitian 
operators and states as state vectors. Even if for convenience, or out of habit, we still use the associated ``old-fashioned'' language from time to time, these 
traditional notions are strictly speaking no longer required. 

As mentioned, in what follows we use an index notation in Hilbert space calculations. 
We model a typical quantum system with the introduction of a finite dimensional complex vector space ${\cal H}^{\alpha}$ ($\alpha = 1, 2, \dots, n$) of 
some specified dimension $n \geq 2$, equipped with a complex conjugation operation $\bf C$, which maps elements of 
${\cal H}^{\alpha}$ to the associated dual space ${\cal H}_{\alpha}$. Thus if $X^{\alpha}$ is a typical element of 
${\cal H}^{\alpha}$, then under the complex conjugation map $X^{\alpha}$ gets mapped by $\bf C$ to the dual vector 
$\bar X_{\alpha}$ in ${\cal H}_{\alpha}$. For any two elements $X^{\alpha}$ and $Y^{\alpha}$ in ${\cal H}^{\alpha}$, if we 
form the sum $a X^{\alpha} + b Y^{\alpha}$, where $a$ and $b$ are complex numbers, then under 
complex conjugation this gets mapped in an antilinear way to $\bar a \bar X_{\alpha} + \bar b \bar Y_{\alpha}$. The 
complex conjugation map acts in such a way that it likewise sends elements of ${\cal H}_{\alpha}$ to ${\cal H}^{\alpha}$, 
with $\bf C^2 = 1$. This structure is to be regarded as in place when we state that ${\cal H}^{\alpha}$ is a finite-dimensional complex 
Hilbert space. 

A compelling case for use of the index notation  is made in Geroch (1974). One could in 
principle use the notation of bras and kets, or an abstract notation; but for our purpose the index notation is more effective, since tensorial objects 
arise frequently in the discussion of quantum systems. The use of an index label on the Hilbert space itself acts as a 
reminder of the structure of the space, a convention that is handy when complicated tensor products of such spaces are 
brought into play. Thus we write 
${\cal H}^{\alpha}$ for the Hilbert space, ${\cal H}_{\alpha}$ for its dual, and
\begin{eqnarray}
{\cal H}^{\alpha}_{\beta} = {\cal H}^{\alpha}  \otimes  {\cal H}_{\beta}
\end{eqnarray} 
for the tensor product of ${\cal H}^{\alpha}$ and ${\cal H}_{\beta}$. Similarly we write
\begin{eqnarray}
{\cal H}^{\alpha \beta} = {\cal H}^{\alpha} \otimes {\cal H}^{\beta}, \quad
{\cal H}_{\alpha \beta} = {\cal H}_{\alpha} \otimes {\cal H}_{\beta}, \quad
{\cal H}^{\alpha \gamma}_{\beta \delta} = {\cal H}^{\alpha} \otimes
{\cal H}_{\beta} \otimes {\cal H}^{\gamma} \otimes {\cal H}_{\delta},
\end{eqnarray} 
and so on.
We use the summation convention to write
$X^{\alpha} \bar X_{\alpha}$ for the  inner product between the ket $X^{\alpha}$ and the bra $\bar X_{\alpha}$, and we say 
that the vector $X^{\alpha}$ is normalised if $X^{\alpha} \bar X_{\alpha} = 1$.  The scheme that we have in mind for the 
manipulation of multi-index tensorial objects in the present setting is essentially the same as the general ``abstract index'' 
setup outlined, for example, in section 3 of Penrose (1968), and sections 2.1-2.2 of Penrose \& Rindler (1984). 
Following the conventional terminology we often refer to the elements of ${\cal H}^{\alpha}$ as ``state vectors", and the 
associated rays as ``states''. This way of speaking is a convenient relic of the old-fashioned approach. 

Now we are in a position to introduce  the ideas of states as positive operators and operations as maps from states 
to states. There is a substantial body of literature concerned with operational approaches to quantum mechanics, much of 
which is relevant to  the ``modern'' approach to the subject. See, for example,  Davies 
(1976), Davies \& Lewis (1970), Haag \& Kastler (1964), Holevo (1982), Krauss (1971, 1983), Mielnik (1968, 1969, 1974), Segal (1947), and references cited therein. 
We begin with the notion of a state. The elements of ${\cal H}^{\alpha}_{\beta}$ act as linear operators on the Hilbert space and its dual. Under complex 
conjugation, any element $F^{\alpha}_{\beta} \in {\cal H}^{\alpha}_{\beta}$ is mapped to an element
$\bar F^{\beta}_{\alpha} \in {\cal H}^{\beta}_{\alpha}$. 
Since the  vector spaces ${\cal H}^{\alpha}_{\beta}$ and ${\cal H}^{\beta}_{\alpha}$ are naturally isomorphic, we can say that $F^{\alpha}_{\beta} \in {\cal H}^{\alpha}_{\beta}$ is mapped under complex conjugation to another element $\bar F^{\alpha}_{\beta} \in {\cal H}^{\alpha}_{\beta}$. 
Then we say that $F^{\alpha}_{\beta} $ is Hermitian if $F^{\alpha}_{\beta} = \bar F^{\alpha}_{\beta} $. 
We shall write ${\cal O}^{\alpha}_{\beta}$ for the space of Hermitian operators. 
An element $F^{\alpha}_{\beta} \in {\cal O}^{\alpha}_{\beta}$ is said to be positive (nonnegative) if $F^{\alpha}_{\beta} \xi^{\beta} {\bar \xi}_{\alpha} \geq 0$ for all $\xi^{\alpha} \in {\cal H}^{\alpha}$, and strictly positive if 
$F^{\alpha}_{\beta} \xi^{\beta} {\bar \xi}_{\alpha} > 0$ 
for all $\xi^{\alpha} \in {\cal H}^{\alpha}$. 
A Hermitian operator is positive if and only if there exists a collection of one or more linearly independent vectors $Z^{\alpha}_i$, with complex conjugates  $\bar Z_{i\alpha}$, each normalised to unity, such that $F^{\alpha}_{\beta} $ is of the form 
\begin{eqnarray}
F^{\alpha}_{\beta} = \sum_{i} f_i Z^{\alpha}_i \bar Z_{i\beta}
\end{eqnarray}
where the $f_i$ are positive constants. 
If a Hermitian operator is strictly positive, then one can find a set of $n$ mutually orthogonal vectors, each normalised to unity, such that $F^{\alpha}_{\beta}$ can be written in the form (1) and where the $f_i$ are strictly positive. 

We shall write 
${^+\cal O}^{\alpha}_{\beta}$ for the cone of positive Hermitian operators. By a ``state" we mean any nontrivial element of ${^+\cal O}^{\alpha}_{\beta}$.  If $A^{\alpha}_{\beta}$ 
and $B^{\alpha}_{\beta}$ are elements of ${^+\cal O}^{\alpha}_{\beta}$, and if $a$ and $b$ are positive constants, not both zero, then $a A^{\alpha}_{\beta} +b B^{\alpha}_{\beta}$ is also an element of  
${^+\cal O}^{\alpha}_{\beta}$. By a ``pure" state, we mean a state of the form $Z^{\alpha} \bar Z_{\beta}$ for some (nontrivial) vector $Z^{\alpha}$. A state 
$F^{\alpha}_{\beta}$ is said to be normalised if $F^{\alpha}_{\alpha}= 1$. In what follows, we do not require that states are necessarily normalised. The terms ``state" and ``density matrix" are used more or less interchangeably in the modern literature; we shall usually reserve the term ``density matrix" for a normalised state. 
It should be clear that with each state vector $X^{\alpha}$ one can associate a corresponding pure state $X^{\alpha} \bar X_{\beta}$. For this reason, it is indeed consistent, even if somewhat misleading, to regard state vectors as  representing a class of ``states". 
In fact, the physics literature seems to be divided on the issue of what constitutes a state. Some physicists take the view, in effect, that an individual quantum system is always in a pure state, and that so-called mixed states, represented by density matrices, correspond to ``ensembles" of pure states (see Hughston {\it et al.}~1993 on this point). Other physicists seem to be happy with the idea that an individual quantum system can be in a mixed state, but that this represents a state of ignorance concerning the ``true'' state of the system, which is pure.  

In our scheme the state of an individual system is represented by a density matrix, which may or may not be pure, and the question of how we use this density matrix, and what calculations we perform with it, depends on the particular model we are constructing. In some models, for example, it can be fruitful to introduce the idea of an ensemble in the form of a probability measure on the space of pure states, which in turn can be represented by a probability measure on the complex projective space $\mathbb {CP}^{n-1}$ associated with the given $n$-dimensional Hilbert space. 

By an ``operation" on a quantum system we mean a positive linear map from ${^+\cal O}^{\alpha}_{\beta}$ to itself. 
Thus we need to consider elements of the space ${\cal O}^{\alpha \,\beta'}_{\alpha' \beta}$ and characterise those elements that constitute positive maps, i.e.~maps from states to states.  
Here we write ${\cal O}^{\alpha \,\beta'}_{\alpha' \beta}$ for the space of Hermitian elements of the tensor product space ${\cal H}^{\alpha \,\beta'}_{\alpha' \beta}$. The primed letters $\alpha'$, $\beta'$, and so on, are regarded as extra letters of the alphabet. 
The action of $\phi^{\alpha \,\beta'}_{\alpha' \beta} \in {\cal O}^{\alpha \,\beta'}_{\alpha' \beta}$ on a state 
$F^{\alpha}_{\alpha'}$ is given by
\begin{eqnarray}
F^{\alpha}_{\alpha'} \rightarrow \phi^{\alpha \,\beta'}_{\alpha' \beta} F^{\beta}_{\beta'} \, . 
\label{operation}
\end{eqnarray}
We say that $\phi^{\alpha \,\beta'}_{\alpha' \beta}$ is a positive map  if under the action indicated above it maps any positive operator to another positive operator. 

\begin{prop}
\label{Positive maps}
The map $\phi^{\alpha \,\beta'}_{\alpha' \beta} \in {\cal O}^{\alpha \,\beta'}_{\alpha' \beta}$ is positive if and only if for 
all $X^{\alpha}$, $Y^{\alpha}$ $\in {\cal H}^{\alpha}$ we have
\begin{eqnarray}
 X^{a'} \bar X_{\alpha} \phi^{\alpha \,\beta'}_{\alpha' \beta} Y^{\beta} \bar Y_{\beta'} \geq 0.
 \label{biquadratic form}
\end{eqnarray}
\end{prop}
\noindent \textit{Proof}. We require that for any positive operator $F^{\beta}_{\beta'}$ the transformed operator 
$\phi^{\alpha \,\beta'}_{\alpha' \beta} F^{\beta}_{\beta'}$ should be positive. 
Thus for all $X^{\alpha}$ we require that
 \begin{eqnarray}
 X^{\alpha'} \bar X_{\alpha} \phi^{\alpha \,\beta'}_{\alpha' \beta} F^{\beta}_{\beta'} \geq 0. 
 \label{operation inequality}
 \end{eqnarray}
In particular, if  $F^{\beta}_{\beta'}$ is a pure state $Y^{\beta} \bar Y_{\beta'}$ then we obtain (\ref{biquadratic form}). Conversely, suppose that (\ref{biquadratic form}) holds. 
Now, any state can be represented as a positively weighted sum of pure states. 
If we fix $X^{\alpha}$ and consider the inequality (\ref{biquadratic form}) for various choices of  $Y^{\alpha}$, we deduce that  (\ref{operation inequality}) holds for any positive $F^{\beta}_{\beta'}$. 
Since this is true for any choice of $X^{\alpha}$, we deduce that $\phi^{\alpha \,\beta'}_{\alpha' \beta}$ is a positive map. \hfill$\Box$
\\\\
\indent Let us write ${^+\cal O}^{\alpha \,\beta'}_{\alpha' \beta}$ for the space of such positive maps. 
For applications we frequently require a stronger condition that limits the class of admissible maps to a subspace of the space of positive maps consisting of so-called ``completely positive" maps. 
The condition of complete positivity ensures that if $\phi^{\alpha \,\beta'}_{\alpha' \beta}$ acts ``locally" on any state of the larger composite system obtained by forming the tensor product of the Hilbert space ${\cal H}^{\alpha}$ and any  ``ancilla" Hilbert space ${\cal H}^{j}$ (not necessarily of the same dimension), then the result is a positive operator. 
More precisely,  suppose that a typical Hilbert space vector of such a composite system is given by $X^{\alpha j} \in {\cal H}^{\alpha j}$. 
One can think of the index ``clump" $\alpha j$ as constituting an index for the Hilbert space vector of the composite system. 
A general state of the composite system is given by a positive operator of the form $F^{\alpha j}_{\beta k}$. 
By saying that $F^{\alpha j}_{\beta k}$ is a positive operator in this situation we mean that for any composite state vector $X^{\beta k}$ we have 
 \begin{eqnarray}
X^{\beta k} \bar X_{\alpha j} F^{\alpha j}_{\beta k}  \geq 0 . 
 \end{eqnarray}
Now clearly if we let $\phi^{\alpha \,\beta'}_{\alpha' \beta}$ operate ``only on the first element of the composite" through the transformation
\begin{eqnarray}
F^{\alpha\, j}_{\alpha'j'} \rightarrow \phi^{\alpha \,\beta'}_{\alpha' \beta} F^{\beta\, j}_{\beta'j'},
\label{transformed state}
\end{eqnarray}
then this gives us a linear map from Hermitian operators to Hermitian operators on the composite system, and we can call $\phi^{\alpha \,\beta'}_{\alpha' \beta}$ a ``local" operation. 
We say that $\phi^{\alpha \,\beta'}_{\alpha' \beta}$ is ``completely positive" if for any such composite system the given transformation takes states to states. 
Note that we do not require here that such an operation should preserve the normalisation of a state. 

\begin{prop}
\label{Completely positive maps}
The map $\phi^{\alpha \,\beta'}_{\alpha' \beta}$ is completely positive if and only if for some $N \geq 1$ there exists a family of operators $K^{\alpha}_{\beta}(i)$ for $i = 1, \dots, N$ such that 
\begin{eqnarray}
 \phi^{\alpha \,\beta'}_{\alpha' \beta} = \sum_{i = 1}^N K^{\alpha}_{\beta}(i) \bar K^{\beta'}_{\alpha'}(i).
 \label{completely positive map}
\end{eqnarray}
\label{prop:2}
\end{prop}

\noindent \textit{Proof}. We require that the transformed state  (\ref{transformed state}) is positive for any initial composite state.  This must be true in particular for a pure state of the form 
$F^{\beta\,j}_{\beta'j'} = \xi ^{\beta\,j} \bar \xi_{\beta'j'}$. The transformed state in that case is
$\phi^{\alpha \,\beta'}_{\alpha' \beta} \xi ^{\beta\,j} \bar \xi_{\beta'j'}$, and  to ensure that it is positive we require that for any composite state vector $X^{\alpha i}$ we have 
\begin{eqnarray}
X^{\alpha'j'} \bar X_{\alpha j} \,\phi^{\alpha \,\beta'}_{\alpha' \beta} \, \xi ^{\beta\,j} \bar \xi_{\beta'j'} \geq 0.
\end{eqnarray}
If we set 
\begin{eqnarray}
Z^{\alpha'}_{\beta'} = X^{\alpha'j'} \bar \xi_{\beta'j'}, \quad \bar Z_{\alpha}^{\beta} = \bar X_{\alpha j} \xi^{\beta j},
\end{eqnarray}
then the inequality (8) takes the simple form 
\begin{eqnarray}
\phi^{\alpha \,\beta'}_{\alpha' \beta} Z^{\alpha'}_{\beta'}  \bar Z_{\alpha}^{\beta} \geq 0.
\end{eqnarray}
This is the condition that the Hermitian form obtained by clumping the indices 
on $Z^{\alpha'}_{\beta'}$ and  $\bar Z_{\alpha}^{\beta}$ should be positive. As a consequence one sees by the theory of positive Hermitian forms that 
$\phi^{\alpha \,\beta'}_{\alpha' \beta}$ admits an expansion of the form (\ref{completely positive map}). Conversely, if  $\phi^{\alpha \,\beta'}_{\alpha' \beta}$ 
takes the form  (\ref{completely positive map}), then it is straightforward to check that the resulting map is completely positive. One needs to verify that for any composite state $F^{\beta\,j}_{\beta'j'}$ and any composite vector $X^{\alpha'j'}$ it holds that
\begin{eqnarray}
X^{\alpha'j'}  \bar X_{\alpha j} \sum_{i = 1}^N K^{\alpha}_{\beta}(i) \,\bar K^{\beta'}_{\alpha'}(i) \,F^{\beta\,j}_{\beta'j'} \geq 0.
\end{eqnarray}
But this follows at once since
\begin{eqnarray}
X^{\alpha'j'}  \bar X_{\alpha j} \sum_{i = 1}^N K^{\alpha}_{\beta}(i) \,\bar K^{\beta'}_{\alpha'}(i) \,F^{\beta\,j}_{\beta'j'} =
\sum_{i = 1}^N W^{\beta'j'} (i) \bar W_{\beta j}(i) \,F^{\beta \,j}_{\beta' j'},
\end{eqnarray}
where for each value of $i$ we define
\begin{eqnarray}
W^{\beta'j'} (i) = X^{\alpha'j'} \bar K^{\beta'}_{\alpha'}(i), \quad \bar W_{\beta j} (i) = \bar X_{\alpha j} K_{\beta}^{\alpha}(i),
\end{eqnarray}
and we thus observe that each term on the right hand side of (12) is positive by virtue of the assumed positivity of the state $F^{\beta \,j}_{\beta'j'}$. 
\hfill $\Box$
\\\\
\indent The result of Proposition~\ref{prop:2} is  the theorem of Choi (1975), with a proof that is on 
account of the use of the index notation perhaps more transparent than the original.

\section{Experiments and state transformations}
\label{Experiments and state transformations} 

Rather than taking a ``one size fits all" approach to quantum theory, the idea is to construct a number of different models. 
Each model involves the specification of a quantum system, the experiments that can be made on it, the possible outcomes, and the resulting state transformations. 
The quantum system  is represented by a Hilbert space ${\mathcal H}^{\alpha}$ along with a state $w^{\alpha}_{\beta}$. 
Each experiment that can be performed  is described by a measurable space $({\Omega},{\mathcal F})$ endowed with some structure that relates it to the Hilbert space 
${\mathcal H}^{\alpha}$ and the state $w^{\alpha}_{\beta}$. 
Here the set ${\Omega}$ represents all of the possible outcomes of chance when the experiment is performed, and ${\mathcal F}$ is a collection of subsets of ${\Omega}$ forming a $\sigma$-algebra. 

More precisely,  we require that ${\Omega}$ itself should belong to ${\mathcal F}$, that the empty set 
$\varnothing$  should belong to ${\mathcal F}$, and that the union of any countable collection of elements of 
${\mathcal F}$ should belong to ${\mathcal F}$. 
Distinct $\sigma$-algebras correspond to distinct experiments. If ${\mathcal E}$ and ${\mathcal F}$ are $\sigma$-algebras on ${\Omega}$, and if ${\mathcal E}$ is a sub-$\sigma$-algebra of ${\mathcal F}$, then we can say that the experiment ${\mathcal F}$ is a refinement of the experiment ${\mathcal E}$. Conversely, 
we can say that the experiment ${\mathcal E}$ is a ``course-grained" version of the experiment 
${\mathcal F}$. In this way one obtains hierarchies of experiments. 

In a given experiment $({\Omega},{\mathcal F})$, if $\omega\in{\Omega}$ is the outcome of chance, then the result of the experiment is the smallest element $A \in{\mathcal F}$ such that $\omega \in A$. We say that $A$ is the smallest element of $ {\mathcal F}$ containing  $\omega$ if $\omega$ belongs to no proper subset of $A$ which is also an element of $ {\mathcal F}$.  In general,  distinct outcomes of chance can give rise to the same result for an experiment. 
It may be that for each $\omega\in{\Omega}$ the subset $\{\omega\} 
\subset {\Omega}$ that only contains $\omega$ belongs to ${\mathcal F}$. 
This happens for example in  ``refined" experiments where the result of the experiment is sufficient to determine the outcome of chance. 
We distinguish between the outcomes of chance (which are elements of the set ${\Omega}$) and the results of experiments (which are minimal elements of 
${\mathcal F}$). Each outcome of chance belongs to a unique minimal element of 
${\mathcal F}$. One might ask why we introduce the entire 
$\sigma$-algebra
 ${\mathcal F}$ if only the ``minimal" elements of ${\mathcal F}$ count as possible results of the experiment represented by ${\mathcal F}$. The reason is that when we consider an experiment, we generally like to consider alongside it the various course-grained versions of the experiment. Thus given ${\mathcal F}$ we wish to consider as well the various $\sigma$-subalgebras associated with it, the minimal elements of which are not necessarily minimal elements of  ${\mathcal F}$.

The next ingredient that we require for the specification of an experiment is a system of state transformations 
${\mathbb T}=\{T(A), \, A\in{\mathcal F}\}$. Thus ${\mathbb T}$ takes the form of a transformation-valued measure on $({\Omega},{\mathcal F})$,  satisfying the following conditions:
\begin{itemize}
\item[(i)] For each $A\in{\mathcal F}$, the associated state transformation $T(A)$ is given by a completely positive map of the form
\begin{eqnarray}
T(A): \, w^{\alpha}_{\beta} \to T^{\alpha \beta'}_{\beta \alpha'}(A) \, w^{\alpha'}_{\beta'} \, .
\end{eqnarray}
\item[(ii)] The system ${\mathbb T}=\{T(A), \, A\in{\mathcal F}\}$ is countably additive. Thus, 
$T^{\alpha \beta'}_{\beta \alpha'}(\varnothing)=0$, and if the sets $\{A_n: \, n\in{\mathds N}\}$ are 
disjoint, and such that $A=\cup_nA_n$, then 
\begin{eqnarray}
T^{\alpha \beta'}_{\beta \alpha'}(A) = \sum_{n \in \mathds N } T^{\alpha \beta'}_{\beta \alpha'}(A_n) \, .
\end{eqnarray}
\item[(iii)] For each $A\in{\mathcal F}$, $T(A)$ is trace-reducing. Thus, we have 
\begin{eqnarray}
\frac{T^{\gamma \beta'}_{\gamma \alpha'}(A) \ w^{\alpha'}_{\beta'}}{w^{\gamma}_{\gamma}} \ \leq \ 1 \, .
\end{eqnarray}
\item[(iv)] $T({\Omega})$ satisfies the law of total probability. Thus, we have
\begin{eqnarray}
\frac{T^{\gamma \beta'}_{\gamma \alpha'}({\Omega}) \ w^{\alpha'}_{\beta'}}{w^{\gamma}_{\gamma}} =1 \, .
\end{eqnarray}
\end{itemize}
Again, the transformations are specified for each element of ${\mathcal F}$, not merely for the minimal elements. In this way, we also determine the relevant transformations for each course-grained version of the experiment. 

Once we have specified the system of transformations on ${\mathcal F}$, then in the experiment $({\Omega},{\mathcal F},{\mathbb T})$, or coursed-grained version thereof, the probability that the outcome of chance $\omega$ lies in the set $A\in{\mathcal F}$ is given by
\begin{eqnarray}
{\mathbb P}(\omega\in A) = 
\frac{T^{\gamma \beta'}_{\gamma \alpha'}(A) \ w^{\alpha'}_{\beta'}}{w^{\gamma}_{\gamma}} \, .
\end{eqnarray}
If $A_\omega$ denotes the smallest element of ${\mathcal F}$ containing  $\omega$,  and if ${\mathbb P}(\omega\in A_\omega)\neq0$, then the normalised state transformation associated with $\omega$ is  
\begin{eqnarray}
w^{\alpha}_{\beta} \to \frac {T^{\alpha \beta'}_{\beta \alpha'}(A_\omega) \ w^{\alpha'}_{\beta'}} {T^{\gamma \beta'}_{\gamma \alpha'}(A_\omega) \ w^{\alpha'}_{\beta'}}\, .
\end{eqnarray}
The state transformations associated with the usual projective measurements (with or without selection) in quantum mechanics take 
this form, and so do the transformations associated with more general discrete POVMs. 
For example, in the maximally course-grained experiment corresponding to a non-selective projective measurement we have 
${\mathcal F} = \{ {\Omega}, \varnothing \}$. 
In that case, the outcome of the measurement is trivial in the sense that we have ${\mathbb P} (\omega \in {\Omega})=1$ and ${\mathbb P} (\omega \in \varnothing)=0$. Nevertheless, the state transformation will in general be nontrivial. For example, in the case of a non-selective projective measurement of the energy of a finite-dimensional system with a nondegenerate Hamiltonian,  no ``outcome" is recorded other than the fact the experiment was done (``the result lies in the admissible set of possible outcomes"), yet the state transforms from a general state to a state which is diagonal in the energy basis. 

In the continuous case, the probability of any particular outcome of chance 
is zero. In that situation we model the state transformations as follows. 
We suppose that there exists a measure $\mu(\rd\omega)$ on ${\Omega}$ and a 
transformation density $t^{\alpha \beta'}_{\beta \alpha'}(\omega)$ with the property that for any $A\in 
{\mathcal F}$ we have 
\begin{eqnarray}
T^{\alpha \beta'}_{\beta \alpha'}(A) = \int_{\Omega} {\mathds 1}\{\omega\in A\} 
t^{\alpha \beta'}_{\beta \alpha'}(\omega) \ \mu(\rd \omega) \, .
\end{eqnarray}
Then the probability distribution for the outcome of chance is 
\begin{eqnarray}
{\mathbb P}(\omega\in \rd\omega) = 
t^{\gamma \beta'}_{\gamma \alpha'} (\omega) \ w^{\alpha'}_{\beta'}
\ \mu(\rd\omega) \, , 
\end{eqnarray}
and the normalised state transformation is given by 
\begin{eqnarray}
w^{\alpha}_{\beta} \to \frac {t^{\alpha \beta'}_{\beta \alpha'}(\omega) 
\ w^{\alpha'}_{\beta'}} {t^{\gamma \beta'}_{\gamma \alpha'}(\omega) \ w^{\alpha'}_{\beta'}}\, .
\end{eqnarray}
In what follows, we shall be concerned mainly with the continuous situation where we have a finite dimensional Hilbert space and the outcome space  ${\Omega}$ has the structure of a manifold on which a natural candidate for the measure $\mu(\rd \omega)$ on  
${\Omega}$ is available.

\section{Tomographic measurements}
\label{Tomographic measurements} 

We consider the case where the measurable space ${\Omega}$ representing the possible outcomes of chance is the manifold ${\mathds C}{\mathbb P}^{n-1}$, the space of pure states associated with the given Hilbert space 
${\mathcal H}^{\alpha}$. This is a rather natural choice to look at first since it does not involve the introduction of any additional structure on the quantum system. Thus all finite dimensional quantum systems admit a version of the following measurement operation.
Write ${{\Omega}} = {\mathds C}{\mathbb P}^{n-1}$, let $x$ denote a typical point in ${{\Omega}}$, and let  $Z^{\alpha}(x)$ denote a representative vector in ${\mathcal H}^{\alpha} \backslash \{0\}$ lying on the fibre above the point $x \in {\Omega}$. 
Then we can construct a system of transformations ${\mathbb T}$ by setting 
\begin{eqnarray}
T^{\alpha \beta'}_{\beta \alpha'}(A) =n \int_{{\Omega}} {\mathds 1}\{x\in A\} 
\frac{Z^{\alpha}(x) Z^{\beta'}(x){\bar Z}_{\beta}(x){\bar Z}_{\alpha'}(x)}
{(Z^{\gamma}(x){\bar Z}_{\gamma}(x))^2} \mu(\rd x) 
\end{eqnarray}
for any element  $A$  of the Borel $\sigma$-algebra on ${\Omega}$. 
Here 
\begin{eqnarray}
\mu(\rd x) = \frac{{\cal D}Z(x)}{\int_{{\Omega}}{\cal D}Z(x)}
\end{eqnarray}
defines the uniform probability measure on ${\mathds C}{\mathbb P}^{n-1}$, where 
\begin{eqnarray}
{\cal D}Z = \frac { \epsilon_{\alpha \beta \cdots \gamma} 
Z^{\alpha} \rd Z^{\beta} \cdots \rd Z^{\gamma} \,
\epsilon^{\alpha \beta \cdots \gamma} {\bar Z}_{\alpha} \rd {\bar Z}_{\beta} \cdots \rd 
{\bar Z}_{\gamma} } { \ (Z^{\gamma}{\bar Z}_{\gamma})^n } \, .
\end{eqnarray}
Clearly,  we have $\mu({\Omega}) = 1$. 
The associated transformation density is then given by 
\begin{eqnarray} 
 t^{\alpha \beta'}_{\beta \alpha'}(x) = n
 \frac{Z^{\alpha}(x) Z^{\beta'}(x){\bar Z}_{\beta}(x){\bar Z}_{\alpha'}(x)}
{(Z^{\gamma}(x){\bar Z}_{\gamma}(x))^2} \, .
\end{eqnarray}
The outcome of chance in such a measurement is a pure state. 
If the initial state is $w^{\alpha}_{\beta}$ and if the outcome of chance is the point $x$ in ${\mathds C}{\mathbb P}^{n-1}$,  the resulting normalised state transformation is 
\begin{eqnarray}
w^{\alpha}_{\beta} \to \frac {Z^{\alpha}(x) {\bar Z}_{\beta}(x)} {Z^{\gamma}(x){\bar Z}_{\gamma}(x)} \, .
\end{eqnarray}
The probability that the outcome lies in a given Borel set $A$ in the space of pure states is 
\begin{eqnarray}
{\mathbb P}(x \in A) = E^{\alpha}_{\beta}(A) \, w^{\beta}_{\alpha} \, , 
\end{eqnarray}
where the ``effect'' $E^{\alpha}_{\beta}(A)$ associated with the set $A$ is given by
\begin{eqnarray}
E^{\alpha}_{\beta}(A)= n\int_{{\Omega}} {\mathds 1}\{x\in A\} 
\frac {Z^{\alpha}(x) {\bar Z}_{\beta}(x)} {Z^{\gamma}(x){\bar Z}_{\gamma}(x)}  \ \mu(\rd x) \, .
\end{eqnarray}
Note that  $E^{\alpha}_{\beta}({\mathit{\Omega}})=\delta^{\alpha}_{\beta}$. 
It follows in particular  that the function $\rho\, : \, {\mathds C}{\mathbb P}^{n-1} \rightarrow 
{\mathds R}^+$ defined by
\begin{eqnarray}
\rho(x) = n \frac{Z^{\alpha}(x) \, w^{\beta}_{\alpha} \, {\bar Z}_{\beta}(x) }
{Z^c(x) {\bar Z}_c(x)} 
\label{rho x}
\end{eqnarray}
is the probability density for the outcome $x \in \rd x$. Thus we have 
\begin{eqnarray}
{\mathbb P}(x \in\rd x) = \rho(x)\ \mu(\rd x). 
\end{eqnarray}
The significance of the factor of $n$ in the expressions above is that it ensures that if the initial state is of the ``totally unbiased'' form
\begin{eqnarray}
w^{\alpha}_{\beta} = \frac{1}{n} \delta^{\alpha}_{\beta}\, ,
\end{eqnarray}
then the probability density is uniform, and we have $\rho(x) = 1$.

Now suppose we consider the situation where we have a large number of independent identical copies of the system, and we make a measurement of this type on each copy. If we analyse the statistics of the measurements, then we can to a good degree of accuracy determine $\rho(x)$, and hence determine the original state $w^{\alpha}_{\beta}$\,. 
More precisely, the ``ensemble'' of measurement outcomes has the density $\rho(x)$, and therefore 
the state $r^{\alpha}_{\beta}$ of the ensemble representing the outcomes  of the measurements is
 \begin{eqnarray}
r^{\alpha}_{\beta} = \int_{ {\Omega}} \rho(x) 
\frac{Z^{\alpha}(x) {\bar Z}_{\beta}(x)}{Z^{\gamma}(x){\bar Z}_{\gamma}(x)} \ \mu(\rd x) .
\end{eqnarray}
The integral can be worked out explicitly by use of the following identity: 
\begin{eqnarray}
 \int_{{\Omega}} 
\frac{Z^{\alpha}(x) \, Z^{\beta'}(x) \,{\bar Z}_{\beta}(x) \,{\bar Z}_{\alpha'}(x)}
{(Z^{\gamma}(x) \,{\bar Z}_{\gamma}(x))^2} \ \mu(\rd x)  = \frac{1}{n(n+1)} 
\big( \delta^{\alpha}_{\beta} \, \delta^{\beta'}_{\alpha'} + \delta^{\alpha}_{\alpha'} \,
\delta^{\beta'}_{\beta}   \big) .  
\label{quadratic delta identity}
\end{eqnarray}
A calculation making use of (\ref{rho x}) and (\ref{quadratic delta identity}) then shows that 
\begin{eqnarray}
r^\alpha_\beta = \frac{1}{n+1} \big( \delta^\alpha_\beta + w^\alpha_\beta \big) \ . 
\end{eqnarray}
We see that the original state $w^\alpha_\beta$ is in general ``diluted'' as a consequence of the measurement operation. But if the initial state is unbiased, then so is the final state. 
In all cases, nevertheless, we can recover the original state from the statistics of the measurement observations since 
\begin{eqnarray}
w^{\alpha}_{\beta} = (n+1) r^{\alpha}_{\beta}  - \delta^{\alpha}_{\beta}  . 
\end{eqnarray}
We can call such an experiment a ``universal quantum measurement'' (UQM) since it can be 
applied to any finite-dimensional quantum system.
No additional structure is required apart from what is already implicit in the original specification of the system 
${\cal H}^\alpha$.  Evidently, this is possible. One can envisage the construction of a machine with the property that given a sample consisting of a large number of identical ``molecules" of some type, all in the same state, the structure of a typical molecule can be determined. 

\section{Disentangling operations}
\label{Disentangling operations} 

Universal quantum measurements of the type just discussed can form elements of composite operations. 
In that case, we introduce more structure on the Hilbert space, but typically not involving the choice of specific observables. %
An example is as follows. 
Consider the Hilbert space of a pair of qubits. 
The Hilbert space has dimension four, and the associated pure state space is 
${\mathds C}{\mathbb P}^{3}$. 
The space of disentangled pure states is a quadric surface in ${\mathds C}{\mathbb P}^{3}$. 
The quadric is a doubly ruled surface, given by the product of two ${\mathds C}{\mathbb P}^{1}$s. 
Each of the ${\mathds C}{\mathbb P}^{1}$s is endowed with the Fubini-Study measure, so the quadric also has a natural uniform measure on it, given by the product measure.
This gives rise to a class of UQMs that we can call ``disentangling operations". Starting with a general (i.e.~mixed) state of 
the two-qubit system, the outcome of a disentangling operation is a point on the quadric. 
The transformation density is the product of the transformation densities associated with the UQMs attached to the 
individual qubits. 

In more detail we have the following. Write $\mathcal H^{A A'}$ ($A, A' = 1, 2$) for the tensor product of the two qubit spaces $\mathcal H^{A }$ and $\mathcal H^{A'}$. A state vector $\xi^{A A'} \in \mathcal H^{A A'}$ is disentangled if it is of the form $\xi^{A A'} = \alpha^A  \beta^{A'}$. The associated quadric $\mathcal Q$ in ${\mathds C}{\mathbb P}^{3}$ is the locus 
\begin{eqnarray}
 \epsilon_{AB} \ \epsilon_{A'B'}\ \xi^{A A'} \xi^{BB'} = 0,
\end{eqnarray}
where $\epsilon_{AB} = - \epsilon_{BA} $. The resulting uniform measure on the quadric is given 
for $x \in \mathcal Q$ by
\begin{eqnarray}
\mu_{\mathcal Q}(\rd x) =
\mu_{\alpha}(\rd x) \mu_{\beta}(\rd x),
\end{eqnarray}
where
\begin{eqnarray}
\mu_{\alpha}(\rd x) =
\frac {  \epsilon_{AB}\alpha^A (x) \rd \alpha^B (x) \,
\epsilon^{CD} \bar \alpha_C (x) \rd \bar \alpha_D (x) }
{ (\alpha^E (x)   \bar \alpha_E (x))^2 }
\end{eqnarray}
and
\begin{eqnarray}
\mu_{\beta}(\rd x) =
\frac {  \epsilon_{A'B'}\beta^{A'} (x) \rd \beta^{B'} (x) \,
\epsilon^{C'D'} \bar \beta_{C'} (x) \rd \bar \beta_{D'} (x) }
{ (\beta^{E'} (x)   \bar \beta_{E'} (x))^2 }.
\end{eqnarray}
The relevant transformation density has its mass concentrated entirely on the quadric, and is given for 
$x \in \mathcal Q$ by
\begin{eqnarray}
t^{AA' CC'}_{BB'DD'}(x) =  4 \
\frac {
\alpha^A (x) \beta^{A'}(x)  \alpha^C (x)  \beta^{C'}(x) \
\bar \alpha_B (x) \bar \beta_{B'}(x) \bar  \alpha_D (x)  \bar \beta_{D'}(x)
}
{(\alpha^E (x)   \bar \alpha_E (x))^2 \ (\beta^{E'} (x)   \bar \beta_{E'} (x))^2 
}
\end{eqnarray}
If the initial state is a prescribed density matrix of the form 
$w^{BB'}_{AA'}$,  then the probability that the outcome lies in
a measurable region $R \subset \mathcal Q$ is
\begin{eqnarray}
{\mathbb P}(\omega\in R) = 4 \ \int_{x \in R}  w_{AA'}^{BB'}  
 \frac {
\alpha^A (x) \beta^{A'}(x)  \bar \alpha_B (x) \bar \beta_{B'}(x) 
}
{ \alpha^E (x)   \bar \alpha_E (x)  \ \beta^{E'} (x)   \bar \beta_{E'} (x) }
\mu_{\mathcal Q}(\rd x) .
\end{eqnarray}
One checks that if the initial state takes the unbiased form 
\begin{eqnarray}
w_{AA'}^{BB'} = \frac {1}{4}  \delta_A^{B}  \delta_{A'}^{B'} ,
\end{eqnarray}
then the distribution of the outcome is uniform over the quadric. On the other hand, if the initial state is a pure singlet state of the form 
\begin{eqnarray}
w_{AA'}^{BB'} = \frac {1}{2}  \epsilon_{AA'} \, \epsilon^{BB'} ,
\end{eqnarray}
then the outcome probability is highest in regions of the quadric corresponding to EPR-Bohm pairs, and vanishes when the two particles have the same state. 
This type of operation may thus be useful as a model for the decay of a particle composed of spin one-half constituents. 

A similar construction applies for entangled states of many-particle systems in higher dimensions, in which case the relevant outcome space is given by the Segre embedding of the product space of the pure state spaces associated with the constituent systems. Whether one views this experiment as a ``measurement'' or a ``procedure'' is to some extent a matter of taste. 
In any case, the effect of the operation is to disintegrate the system into its constituents.

\section{Coherent measurements}
\label{Coherent measurements} 

Consider a three-dimensional Hilbert space, for which the space of pure states has the structure of the complex projective space ${\mathds C}{\mathbb P}^2$ endowed with the Fubini-Study metric. 
Let ${\cal C}$ be a real conic curve in ${\mathds C}{\mathbb P}^2$. 
By ``real'' we mean the following: we require that for any point $x$ in ${\cal C}$ the associated complex 
conjugate line (representing pure states orthogonal to $x$) is tangent to the conic. 
Such a setup is equivalent to representing $\cal H^\alpha$ as a space of symmetric spinors 
${\cal H}^{AB}$, with a typical element $z^{AB}$ (where $A=1,2$) so $z^{AB}=z^{BA}$. 
The conic is given by 
\begin{eqnarray}
\epsilon_{AB}\, \epsilon_{CD} \, z^{AC}z^{BD}=0 \ ,
\end{eqnarray}
where $\epsilon_{AB} = - \epsilon_{BA}$. The solution to this quadratic equation takes the form 
\begin{eqnarray}
z^{AB} = \phi^A \phi^B 
\end{eqnarray}
for some $\phi^A$. 
The associated complex conjugate line consists of all states $x^{AB}$ such that 
\begin{eqnarray}
{\bar\phi}_A{\bar\phi}_B x^{AB} = 0 . 
\end{eqnarray}
Thus the pure states orthogonal to the point $z^{AB} = \phi^A \phi^B$ on the conic are of the form 
\begin{eqnarray}
x^{AB} = {\bar\phi}^{(A}\alpha^{B)} 
\end{eqnarray}
for some $\alpha^A$. %
But any such state lies on a line tangent to ${\cal C}$, the tangent point being 
${\bar\phi}^{A}{\bar\phi}^{B}$. 
We can use ${\cal C}$ as the outcome space of a special class of measurements. 
For any initial spin-one mixed state $w^{AB}_{CD}$ the outcome of the measurement is a point of the 
conic ${\cal C}$, that is 
to say, a pure spin state with a definite direction for the axis of spin. 
Thus, the state transformation is
\begin{eqnarray}
w^{AB}_{CD} \to \phi^A \phi^B {\bar\phi}_C{\bar\phi}_D \ / \ ( \phi^E {\bar\phi}_E)^2 \ .
\end{eqnarray}
The probability that the outcome lies in a given Borel set $A\subset{\cal C}$ is given by 
\begin{eqnarray}
{\mathbb P}(\omega\in A) = 3 \int_A \frac {w^{AB}_{CD} \ \phi^C (x) \phi^D (x) {\bar\phi}_A (x) {\bar\phi}_B (x)  } 
{( \phi^E (x) {\bar\phi}_E (x) )^{2}} \ \mu(\rd x) , 
\end{eqnarray}
where $\mu(\rd x)$ is the uniform probability measure on ${\cal C}$ induced by the Veronese embedding of 
${\mathds C}{\mathbb P}^1$ in ${\mathds C}{\mathbb P}^2$ as a rational curve. 
Such an experiment on a spin-one system can be interpreted physically as a ``measurement of the direction 
of the axis of the spin'' of the particle. 
The result of the experiment is an answer to the question ``what is the direction of the spin 
axis of the particle?".
The state then transforms from the original state to a pure state, which is the unique state lying on the conic
that has that axis of spin. 
Similar formulae apply for higher spin systems, in which case the defining structure involves a 
rational curve of degree $2s$ in ${\mathds C}{\mathbb P}^{2s}$ (the twisted cubic curve, the rational quartic curve, and so on). See Brody \& Hughston (2001) for a discussion of the geometry of higher spin systems and the role played by rational curves. 

More generally, one obtains a broader class of ``coherent'' measurements based on the Veronese 
embedding of ${\mathds C}{\mathbb P}^{n-1}$ as a rational variety of degree $d$ in 
${\mathds C}{\mathbb P}^{N-1}$, where $N$ is given by  
(\ref{Veronese embedding dimension}). For example, if $n = 3$, we obtain the embedding of 
${\mathds C}{\mathbb P}^{2}$ as a rational surface of degree $d$ in 
${\mathds C}{\mathbb P}^{N-1}$ for $N =  \half (d+1)(d+2)$. These varieties are the 
manifolds of so-called generalised coherent states or SU($n$) coherent states (Brody \& Graefe 2010). 
Hence starting from an arbitrary pure or mixed state of the system, the outcome of the 
measurement results in a coherent state. 

Another class of UQM that one can consider, which we hope to discuss elsewhere, allows for a direct measurement of the ``mean energy" of a quantum system. In this case, the additional structure required is the specification of a Hamiltonian operator. Unlike a standard projective measurement of the energy, the outcome of a mean-energy measurement lies in a continuum of possible values between the highest and lowest eigenvalues of the Hamiltonian. The mean-energy measurement operation is closely linked to the existence of the so-called mean-energy ensemble (Brody \& Hughston 1998, 1999), and may be of some relevance in connection with the quantum thermodynamics of finite systems. 

It is interesting to observe that nearly all of the examples we have considered rely rather heavily, or at least so it seems, on what Mielnik (2001) refers to as the ``Ptolomean structure'' of quantum mechanics---namely, the endless hierarchy of tensor products of Hilbert spaces, along with the various notions of entanglement thus entailed, upon which so many of the modern finite-dimensional applications of the theory appear to rest. 

But it is an open question whether the Ptolomean structure really is an essential part of physics. Can it be softened somewhat, perhaps  in the way in which the rigid Minkowskian geometry of special relativity survives in a weaker sense in the tangent space of a general relativistic space-time? This idea is one of the motivations for the geometric approach to quantum mechanics (see, e.g., Ashtekar \& Schilling 1998, Bengtsson \& ${\dot{\rm Z}}$yczkowski 2006, Brody \& Hughston 2001, Gibbons 1992, Hughston 1995, 1996, Kibble 1979, Mielnik 1968, 1974, and references cited therein). But whereas most of the attempts at generalising quantum theory have focussed either on the consideration of generalised state spaces, or generalisations of the notion of observables as phase space functions, relatively little has been pursued so far in the direction of generalised measurements in the context of a nonlinear theory. It makes sense therefore to consider first those classes of operations that depend only on a minimal amount of structure. 

\end{document}